\documentclass[aps,prc,twocolumn,superscriptaddress,showpacs,floatfix,nofootinbib]{revtex4-1}
\usepackage{amsmath,graphicx}

\begin{document}

\title{Mapping the hydrodynamic response to the initial geometry in heavy-ion collisions}

\author{Fernando G.  Gardim}
\author{Fr\'ed\'erique Grassi}
\affiliation{
Instituto de F\'\i sica, Universidade de S\~ao Paulo, C.P. 66318, 05315-970, S\~ao Paulo-SP, Brazil}
\author{Matthew Luzum}
\author{Jean-Yves Ollitrault}
\affiliation{
CNRS, URA2306, IPhT, Institut de physique th\'eorique de Saclay, F-91191
Gif-sur-Yvette, France}
\date{\today}

\begin{abstract}
We investigate how the initial geometry of a heavy-ion collision is transformed into final flow observables by solving event-by-event ideal hydrodynamics with realistic fluctuating initial conditions.  We study quantitatively to what extent anisotropic flow ($v_n$) is determined by the initial eccentricity $\varepsilon_n$ for a set of realistic simulations, and we discuss which definition of $\varepsilon_n$ gives the best estimator of $v_n$.  We find that the common practice of using an $r^2$ weight in the definition of $\varepsilon_n$ in general results in a poorer predictor of $v_n$ than when using $r^n$ weight, for $n > 2$.   We similarly study the importance of additional properties of the initial state.  For example, we show that in order to correctly predict $v_4$ and $v_5$ for non-central collisions, one must take into account nonlinear terms proportional to $\varepsilon_2^2$ and $\varepsilon_2\varepsilon_3$, respectively.  We find that it makes no difference whether one calculates the eccentricities over a range of rapidity, or in a single slice at $z=0$, nor is it important whether one uses an energy or entropy density weight. This knowledge will be important for making a more direct link between experimental observables and hydrodynamic initial conditions, the latter being poorly constrained at present.
\end{abstract}


\maketitle

\section{Introduction}
\label{s:introduction}
Anisotropic flow~\cite{Voloshin:2008dg} is one of the most important probes of ultrarelativistic nucleus-nucleus collisions.
While early studies~\cite{Ollitrault:1992bk} focused on elliptic flow generated by the almond shape of the interaction region in non-central collisions, most of the recent activity concerns the effect of fluctuations in the initial geometry~\cite{Miller:2003kd}. Such fluctuations result in fluctuations of elliptic flow~\cite{Alver:2006wh}, and also in new types of flow, such as triangular flow~\cite{Alver:2010gr} and higher harmonics. These new flow observables have been recently measured at RHIC~\cite{Adare:2011tg,Sorensen:2011fb} and LHC~\cite{Aamodt:2011vk,Aamodt:2011by,Jia:2011hf,Li:2011mp}.

Flow phenomena are best modeled with ideal~\cite{Huovinen:2006jp}
or viscous~\cite{Romatschke:2009im} hydrodynamics. Event-by-event
hydrodynamics~\cite{Hama:2004rr} provides a natural way of
studying flow fluctuations: one typically supplies a set of
initial conditions, then evolves these initial conditions through
ideal
\cite{Hama:2004rr,Petersen:2008dd,Holopainen:2010gz,Werner:2010aa,Qiu:2011iv}
or viscous~\cite{Schenke:2010rr} hydrodynamics, then computes
particle emission at the end. Observables are finally averaged
over a large number of initial conditions, much in the same way as
they are averaged over events in an actual experiment.

The largest source of uncertainty in these hydrodynamic models is
the initial conditions~\cite{Luzum:2008cw,Heinz:2011kt}---that is,
the state of the system after which it has sufficiently
thermalized or isotropized for the hydrodynamic description to be
valid.  Several models of initial geometry fluctuations have been
proposed~\cite{arXiv:0710.5731,arXiv:0707.0249,arXiv:0805.4411,
arXiv:1108.4862,arXiv:1111.3347}. The usual procedure is to choose 
one or two of these simple models
for the initial conditions and calculate the resulting flow
observables.  Significant progress has been made recently by
simultaneously comparing to several of the newly-measured flow
observables.  With this approach, hydrodynamic calculations can be
used to rule out a particular model of initial conditions if
results do not match experimental data~\cite{Alver:2010dn, Adare:2011tg, Qiu:2011hf}. But it does not tell us
{\it why\/} a particular model fails. In order to constrain the
initial state directly from data, we need to identify which
properties of the initial state determine a given observable.
These constraints can then provide valuable guidance in the
construction of better, more sophisticated models of the
early-time dynamics.

It is well known that elliptic flow is largely determined by the
participant eccentricity~\cite{Alver:2006wh}. Teaney and
Yan~\cite{Teaney:2010vd} have introduced a cumulant expansion of
the initial density profile, in which the participant eccentricity
is only the first term in an infinite series, and they have
suggested that the hydrodynamic response may be improved by adding
higher-order terms, but to our knowledge their suggestion has
never been checked quantitatively. Other expansions have also been
suggested~\cite{Gubser:2010ui}. As for triangular flow, $v_3$,
symmetry considerations have been used to argue that it should be
created by an initial triangularity $\varepsilon_3$, but several
definitions of $\varepsilon_3$ are in use
\cite{Alver:2010gr,Petersen:2010cw} and it has never been
investigated which is a better predictor of $v_3$. Finally, it has
been shown that higher harmonics~\cite{Qiu:2011iv,Gardim:2011re}
$v_4$ and $v_5$ are in general {\it not\/} proportional to the
corresponding $\varepsilon_4$ and $\varepsilon_5$.  A possible
better estimator was recently suggested \cite{Luzum:2011mm}, but
it has not been checked quantitatively.

The goal of this paper is to improve our understanding of the hydrodynamic response to initial fluctuations.
We carry out event-by-event ideal hydrodynamic calculations with realistic initial conditions and then quantitatively compare the final values of $v_n$ with estimates derived from the initial density profile. We are thus able to systematically determine the best estimators of flow observables $v_n$,  $n=$ 2--5 from the initial transverse density profile.

\section{A systematic approach to characterizing the hydrodynamic response}
\label{s:expansion}

In hydrodynamics, the momentum distribution of particles at the end of the evolution is completely determined by initial conditions.
Current models of initial conditions predict the system at early times to consist of flux tubes or other string-like structures that are extended
longitudinally~\cite{Dumitru:2008wn}, with an approximate boost invariance at mid-rapidity. Most models also predict that the initial transverse flow, if any, is
small~\cite{Broniowski:2008vp,Vredevoogd:2008id} (except for possible fluctuations of the initial flow velocity~\cite{arXiv:1108.5535}). Under these approximations, any observable is a completely deterministic functional of the transverse energy density profile $\rho(x,y)$.

This density in turn can be completely characterized by a set of complex moments~\cite{Teaney:2010vd}
\begin{eqnarray}
\label{moments}
W_{p+q,p-q}&\equiv& \int (x+iy)^p(x-iy)^q \rho(x,y) dxdy\cr
&=&\int r^{p+q} {\rm e}^{i(p-q)\phi}\rho(r\cos\phi,r\sin\phi)r{\rm d}r{\rm d}\phi\cr
&\equiv& W_{0,0} \{  r^{p+q} {\rm e}^{i(p-q)\phi} \},
\end{eqnarray}
where $ \{ \cdots \}$ denotes an average value over the transverse plane weighted by $\rho(x,y)$.
Small values of the first index $(p+q)$ correspond to small powers of $|\vec{k}|$ in a two-dimensional Fourier transform of $\rho(x,y)$, and thus describe large-scale structure, while moments with larger values are more sensitive to small-scale structure.  The second index indicates the rotational symmetry of each moment.

If the system has $\phi\to\phi+\pi$ symmetry, all odd moments
(i.e., with $p-q$ odd) vanish. In a symmetric heavy-ion collision,
the symmetry between target and projectile implies an approximate
$\phi\to\phi+\pi$  symmetry in a centered coordinate system
defined by $W_{1,1}=0$ (used throughout this article). This
symmetry is only broken by quantum fluctuations in the
wavefunction of incoming nuclei~\cite{Alver:2010gr}. Therefore,
odd moments are typically small relative to even moments.
Similarly, central heavy-ion collisions are rotationally symmetric
except for fluctuations, so that all moments with $p\not= q$ are
small. For semi-central or peripheral collisions, however, the
interaction area is almond shaped~\cite{Ollitrault:1992bk},
resulting in sizable moments in the 2nd Fourier harmonic $p-q=2$.
For example, the familiar participant eccentricity $\varepsilon_2$
and participant plane $\Phi_2$ are defined by~\cite{Alver:2006wh}
\begin{equation}
\label{defeps2}
\varepsilon_2 {\rm e}^{2i\Phi_2} \equiv -\frac{W_{2,2}}{W_{2,0}}=-\frac{\{r^2 {\rm e}^{2i\phi}\}}{\{r^2\}}.
\end{equation}
The value of $\varepsilon_2$ is typically $0.3$ in a semicentral heavy-ion collision: anisotropies are small, and so it is natural to expect that the hydrodynamic response can be ordered into a Taylor's series.
Higher-order even harmonics are smaller:
The fourth-order anisotropy $\varepsilon_4 \equiv |W_{4,4}|/W_{4,0}$ is typically of order $(\varepsilon_2)^2$~\cite{Lacey:2010hw}, so that it can be treated in practice as a higher order term in a Taylor series expansion.

Any observable can generally be written as a function of moments of
$\rho$. In this paper, we focus on anisotropic flow, $v_n$, which,
along with the event-plane angle $\Psi_n$, is defined by
\begin{equation}
\label{defvn}
v_n {\rm e}^{in\Psi_n}=\left\{  {\rm e}^{in\phi_p}\right \}.
\end{equation}
Here $ \{ \cdots \}$ denotes an average over the distribution of
particle momenta in one event. In hydrodynamics, this is a smooth
(boosted thermal) probability distribution. 
In a real-world collision, $v_n$ and $\Psi_n$ must be inferred from a
finite sample of particles. The resulting statistical error makes an
event-by-event 
determination of $v_n$ impossible in current experiments---only
event-averaged quantities are reliable. In theoretical calculations,
however, we can accurately determine $v_n$ and $\Psi_n$ in every
event in order to study precisely how they depend on initial
conditions. 

The symmetries of $v_n$ restrict what combinations of moments of the initial distribution it can depend on.  For example,
to first order in anisotropies, 
the rotational symmetry of $v_n$ implies that it is a linear combination of  moments in the same harmonic:
\begin{equation}
\label{leadingorder}
v_n {\rm e}^{in\Psi_n}=\sum_{p=0}^{\infty} k_{n+2p,n} W_{n+2p,n},
\end{equation}
where the coefficients $k_{n+2p,n}$ are (dimensionful) functions
of (the infinite set of) rotationally symmetric moments
$W_{2m,0}$. The conventional eccentricity scaling of elliptic
flow, $n=2$, amounts to truncating the series to the first term,
$p=0$.  I.e., the statement that $v_2\propto\varepsilon_2$ is a
statement that the hydrodynamic response is sensitive only to the
large-scale structure of the initial density distribution, with
the response to small-scale structure damped in comparison.  This
statement has not been quantitatively tested until now.

Teaney and Yan~\cite{Teaney:2010vd} have suggested that including a higher-order term $p=1$ in addition to the lowest order may improve the accuracy (actually, they listed cumulants instead of moments, but it is equivalent).
This conjecture will be checked quantitatively in Sec.~\ref{s:results}.

In this work we will use the following notation for the
dimensionless eccentricity $\varepsilon_{m,n}$ and the
corresponding orientation angle $\Phi_{m,n}$ in a given event:
\begin{equation}
\label{epsmn}
\varepsilon_{m,n} {\rm e}^{in\Phi_{m,n}} \equiv - \frac {\{r^m {\rm e}^{in\phi}\}} {\{r^m\}},
\end{equation}
and we use the shorthand notations $\varepsilon_n\equiv\varepsilon_{n,n}$, $\Phi_n\equiv\Phi_{n,n}$.
If $m-n$ is even and positive, the numerator of Eq.~(\ref{epsmn}) is $W_{m,n}/W_{0,0}$.
If $m$ is even, the denominator  is $W_{m,0}/W_{0,0}$.

Gubser and Yarom have proposed a different basis for the expansion~\cite{Gubser:2010ui}, which can be seen as a partial resummation of the infinite series (\ref{leadingorder}).
To first order in anisotropies, they write $v_n {\rm e}^{in\Psi_n}\propto f_n e^{in\Phi^{GY}_n}$, where $f_n$ and $\Phi^{GY}_n$ are solely determined by the initial density profile. The first harmonics are given by
\begin{eqnarray}
\label{GY}
f_1 e^{i\Phi^{GY}_1}&=&-\left\{ \frac{qre^{i\phi}}{1+q^2 r^2}\right\}\cr
f_2 e^{2i\Phi^{GY}_2}&=&-\left\{ \frac{q^2r^2e^{2i\phi}}{1+q^2 r^2}\right\}\cr
f_3 e^{3i\Phi^{GY}_3}&=&-\left\{ \frac{q^3 r^3e^{3i\phi} }{\left(1+q^2 r^2\right)^2}\right\},
\end{eqnarray}
with $q^2\equiv 1/\{r^2\}=W_{0,0}/W_{2,0}$.
Strictly speaking, this expansion scheme is of obvious relevance only for a conformal equation of state and for a particular initial density profile falling more slowly at large $r$ than realistic profiles. Nevertheless, we find it instructive to test how this expansion compares with conventional eccentricity scaling with realistic initial conditions.

Finally, one can also consider moments of the entropy density profile instead of energy density, which we also test in the following.

The goal here is to determine which moment or combination of moments serve as a best estimator for flow observables $v_n$ with $n=$ 2--5.
We shall see in particular that $v_4$ and $v_5$ are not well described by the leading-order expansion (\ref{leadingorder}) and require nonlinear terms, which are also constrained by symmetry.
Note that symmetry considerations alone exclude a linear mixing between the second and third harmonics, as proposed in Ref.~\cite{Qin:2010pf}.

\section{Determining the best estimator of $v_n$}
\label{s:method}

The goal of this work is to test to what extent $v_n$ and $\Psi_n$ are
correlated with quantities derived from the initial transverse density
distribution, such as $\varepsilon_n$ and $\Phi_n$.
Previously, the correlation of anisotropic flow with the initial
geometry has been studied by plotting the distribution of
$\Psi_n-\Phi_n$~\cite{Holopainen:2010gz,Petersen:2010cw,Qiu:2011iv} 
or by displaying a scatter plot of $v_n$ versus
$\varepsilon_n$~\cite{Gardim:2011qn}.  
In this paper, we carry out a global analysis which studies
both aspects simultaneously and quantitatively. 

For a given event, we write
\begin{equation}
\label{whatwedo}
v_n {\rm e}^{in\Psi_n}=k\varepsilon_n  {\rm e}^{in\Phi_n}+{\cal E},
\end{equation}
where $k$ is an unknown proportionality constant. The first term
in the right-hand side defines the estimate for $v_n$ from the
initial eccentricity, and the last term ${\cal E}$ is the
difference between the calculated flow and the proposed estimator,
or the error in the estimate (note that ${\cal E}$ is complex). No
known estimator can perfectly predict $v_n$ in every event (for
example, two events with the same triangularity can be constructed
to have very different triangular
flow~\cite{Gardim:2011re,Andrade:2011fu}).
 The best estimator, then, should be defined as the one that minimizes the mean square error $\langle |{\cal E}|^2\rangle$, where $\langle \cdots \rangle$ denotes an average over  events in a centrality class. A straightforward calculation shows that the best value of $k$ is
 \begin{equation}
\label{bestk}
k=\frac{\left\langle  \varepsilon_n v_n \cos(n(\Psi_n-\Phi_n)) \right\rangle}{\left\langle  \varepsilon_n^2\right\rangle}.
\end{equation}
Inserting Eq.~(\ref{bestk}) into Eq.~(\ref{whatwedo}), one obtains the best estimator of $v_n$ from $\varepsilon_n$ in a centrality class.
Using Eqs.~(\ref{whatwedo}) and (\ref{bestk}), one finally derives the mean-square error
\begin{equation}
\label{rme}
\left\langle |{\cal E}|^2\right\rangle=\left\langle v_n^2\right\rangle-k^2\left\langle\varepsilon_n^2\right\rangle.
\end{equation}
This shows that the rms value of the best estimator, $|k|\sqrt{\left\langle\varepsilon_n^2\right\rangle}$, is always smaller than the rms value of
$v_n$. In the next section, we compute
$k\sqrt{\left\langle\varepsilon_n^2\right\rangle}/\sqrt{\left\langle v_n^2\right\rangle}$ for various definitions of $\varepsilon_n$ and several values of $n$.
The closer the ratio to 1, the better the estimate. Using Eq.~(\ref{rme}), a ratio of $0.95$ corresponds to a rms error of 31\%.
A change of sign in the ratio signals that the estimator is anticorrelated to $v_n$.

According to the discussion in Sec.~\ref{s:expansion}, an improved estimator may be obtained by adding more terms in Eq.~(\ref{whatwedo}), e.g.,
\begin{equation}
\label{more}
v_n {\rm e}^{in\Psi_n}=k\varepsilon_n  {\rm e}^{in\Phi_n}+k'\varepsilon'_n  {\rm e}^{in\Phi'_n}+{\cal E},
\end{equation}
where $\varepsilon'_n$ and $\Phi'_n$ are other quantities determined from the initial density profile (for example, the next higher cumulant). The best estimator is now given by the following system of equations
\begin{eqnarray}
\label{system2}
\langle  \varepsilon_n v_n\cos(n(\Psi_n-\Phi_n)) \rangle&=&k'\langle  \varepsilon_n \varepsilon'_n \cos(n(\Phi'_n-\Phi_n))\rangle \cr
&&+k\langle  \varepsilon_n^2\rangle\cr
\langle  \varepsilon'_n v_n\cos(n(\Psi_n-\Phi'_n))\rangle&=&k\langle  \varepsilon_n \varepsilon'_n \cos(n(\Phi_n-\Phi'_n)) \rangle\cr
&&+k'\langle \varepsilon_n^{\prime 2}\rangle,
\end{eqnarray}
which can be solved for $k$ and $k'$.
The rms value of the best estimator and the rms error are related by an equation analogous to Eq.~(\ref{rme}):
\begin{equation}
\label{rme2}
\left\langle |{\cal E}|^2\right\rangle=\left\langle v_n^2\right\rangle-\left\langle \left| k\varepsilon_n {\rm e}^{in\Phi_n}+k'\varepsilon'_n  {\rm e}^{in\Phi'_n}\right|^2\right\rangle.
\end{equation}
One can show that the rms error is always smaller with two terms, Eq.~(\ref{rme2}), than with only one of the terms, Eq.~(\ref{rme}).
This is intuitive if one thinks
of Eqs.~(\ref{whatwedo}) and (\ref{more}) as fits to $v_n$: adding more parameters improves the quality of the fit.

\section{Results}
\label{s:results}

We simulate Au-Au collisions at the top RHIC energy using
the hydrodynamic code NeXSPheRIO~\cite{Hama:2004rr}.
NeXSPheRIO solves the equations of relativistic ideal hydrodynamics
using initial
conditions provided by the event generator
NeXus~\cite{Drescher:2000ha}.
Fluctuations in initial conditions are studied by
generating 150 NeXus events in each of the 10\%  centrality classes studied,
and solving the equations of ideal hydrodynamics independently for each event.  In addition, 115 NeXus events with zero impact parameter were used in order to study very central collisions.
NeXSPheRIO provides a good description of rapidity and transverse
momentum spectra~\cite{Qian:2007ff},
and elliptic flow $v_2$~\cite{Andrade:2008xh}.
In addition, it reproduces the long-range structures observed in two-particle
correlations~\cite{Takahashi:2009na}.

The code NeXSPheRIO emits particles at the end of the hydrodynamical evolution using a Monte-Carlo generator. Anisotropic flow $v_n$, and the corresponding event-plane angle $\Psi_n$ are defined from Eq.~(\ref{defvn}), where
$\{\cdots\}$ denotes an average over all particles in the pseudorapidity interval $-1<\eta<1$.

This work requires an accurate determination of $v_n$ in each
hydrodynamic event, and so associated with each initial condition, we
generate approximately $6\times 10^5$ particles by computing particle
production with $N$ Monte-Carlos.  This allows for a much better event
plane resolution and much smaller statistical error (the actual
multiplicity in an event is $\approx 6\times 10^5/N$). We compute
$v_2$ to $v_5$. We do not compute $v_1$ because it changes sign as a
function of transverse momentum~\cite{Gardim:2011qn}, 
so that an average with equal weighting such as in Eq.~(\ref{defvn}) is not appropriate~\cite{Luzum:2010fb}.   An analysis of directed flow is left to future work.
The relative statistical errors on $v_2$ to $v_5$ in a given event are $3.7\%$,  $5.7\%$,  $9.8\%$,  $20\%$,  respectively.
The rms error on the event planes $\Psi_2$ to $\Psi_5$ are  $1^\circ$, $1^\circ$, $1.5^\circ$, $2^\circ$.
This means that the event-plane resolution~\cite{Poskanzer:1998yz} for
$v_5$ is as large as 0.98, and even closer to 1 for all other harmonics.

In such a 3-dimensional calculation, there is more than one way to
define the transverse energy density profile that is used as a
weight when calculating the eccentricities in Eq.~(\ref{epsmn}).
We show results obtained by averaging over the transverse energy
density profile at $z=0$ (i.e., central space-time rapidity
$\eta_s=0$).  Though the results are not shown, we have found that
averaging over the space-time rapidity interval $-1<\eta_s<1$
results in predictors of equal quality.

It should be noted that the error from these predictors is likely to be larger in our calculations than in many others, for several reasons.
First, our hydrodynamical calculations are based on NeXus initial conditions which contain (fluctuating) initial flow, as well as longitudinal fluctuations, and so the final flow measured in a particular pseudorapidity window is not entirely determined by the initial transverse geometry.  In addition, there are statistical fluctuations from the finite number of particles generated at the end of the hydro evolution. These issues set a limit on the rms error introduced in Sec.~\ref{s:method}, which cannot go to zero.
A hydrodynamic calculation with less or no initial flow, in 2+1 dimensions,  or that calculates flow from a continuous distribution at freeze out, though perhaps less realistic, will likely result in a smaller error for the same estimator.  Likewise, a non-zero viscosity may cause higher-order cumulants to decrease in importance, improving the predictive power of the lowest moments $\varepsilon_n$.
In this sense, these results represent something of a worst-case scenario.
\subsection{Elliptic flow}
\begin{figure}
\includegraphics[width=\linewidth]{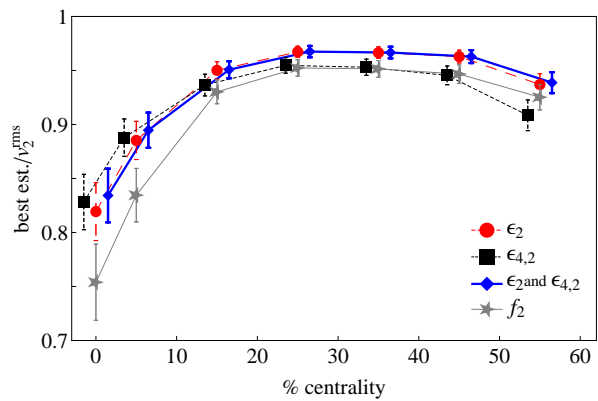}
\caption{(Color online) Best estimator for elliptic flow divided by rms $v_2$ for various combinations of moments of the initial energy density. The leftmost points correspond to 115 events with exactly zero impact parameter.  Error bars represent statistical uncertainty from the finite number of events.  Diamonds (squares) have been given an x-offset of 1.5 (-1.5) for readability.}
\label{fig:v2}
\end{figure}
%
%
Elliptic flow is usually thought of as a hydrodynamic response to the initial eccentricity:
$v_2 {\rm e}^{2i\Psi_2}=k \varepsilon_2 {\rm e}^{2i\Phi_2}$,
where the resulting event plane $\Psi_2$ approximately coincides with the participant plane $\Phi_2$~\cite{Alver:2006wh}.
The right-hand side of this equation defines an estimator of $v_2$.

Fig.~\ref{fig:v2} displays the ratio of the rms value of the best estimator to the rms value of the calculated $v_2$ as a function of centrality.
The proportionality constant $k$ is determined using Eq.~(\ref{bestk}) independently for each centrality class.
For central collisions, where all anisotropies are due to fluctuations, the best estimator is able to reproduce over 80\% of $v_2$. This means that the participant eccentricity correctly captures the physics of $v_2$ fluctuations, but that the event-plane $\Psi_2$ fluctuates around the
participant plane $\Phi_2$~\cite{Holopainen:2010gz,Petersen:2010cw,Qiu:2011iv}, and/or that $v_2$ has sizable fluctuations for a given $\varepsilon_2$. For mid-central collisions, elliptic flow is driven by the almond-shaped overlap area, therefore fluctuations are smaller and the estimate is better.
The value of $k$ slightly decreases with centrality, but very slowly. It is approximately equal to 0.16.  Note that $k$ does not represent the ratio of the magnitudes $v_2/\varepsilon_2$ in a typical collision, and should not be compared with $v_2/\varepsilon_2$ obtained from smooth initial conditions, nor to the ratio of the average elliptic flow to the average eccentricity $v_2/\varepsilon_2$ which are both larger~\cite{Alver:2010dn}.
%

As explained in Sec.~\ref{s:expansion}, the participant
eccentricity is one term out of an infinite series of moments (or cumulants)
allowed by symmetry. There is no fundamental reason why the first
term in Eq.~(\ref{leadingorder}) must be more important than
higher-order terms. In order to check quantitatively this issue,
we define another estimator of $v_2$, corresponding to the term
$p=1$ in Eq.~(\ref{leadingorder}): $v_2 {\rm
e}^{2i\Psi_2}=k\varepsilon_{4,2} {\rm e}^{2i\Phi_{4,2}}$. The
difference with usual participant eccentricity scaling is that
larger values of $r$ are given more weight. As shown in
Fig.~\ref{fig:v2}, this estimate is essentially as good as the
usual participant eccentricity. A closer look reveals that it is
slightly better for central collisions, and slightly worse for
non-central collisions. This means that $v_2$ is driven more by
the periphery of the fireball for central collisions than for
peripheral collisions. The result that, for central collisions,
anisotropic flow  $v_n$ is sensitive to the geometry of the outer
layers of the system is in agreement with
Ref.~\cite{Gardim:2011re,Andrade:2010xy}.
There is a limit, however.  We have checked that $\varepsilon_{6,2}$ gives a worse estimate than $\varepsilon_{4,2}$ for all centralities, indicating that higher-order moments, and thus the extreme periphery, are indeed less important.

Using entropy density instead of energy density to calculate moments also gives a somewhat greater weight to larger $r$.  Indeed we have found that, calculating $\varepsilon_2$ with an entropy density weight gives results (not shown) that are between the results for $\varepsilon_2$ and $\varepsilon_{4,2}$ --- that is, slightly better for central collisions and slightly worse for peripheral collisions, but in general the result is very close to the result for $\varepsilon_2$ calculated using energy density.  In general either predictor appears to be essentially as good as the other.

Next, we test if the quality of the estimator is improved by combining $\varepsilon_2$ and $\varepsilon_{4,2}$, as in Eq.~(\ref{more}). The improvement is marginal, which means that adding the next term in the cumulant expansion~\cite{Teaney:2010vd} does not significantly improve the determination of the event plane from initial conditions. This implies that the small-scale structure of the initial conditions is unlikely to be responsible for the part of elliptic flow that is not explained by $\varepsilon_2$, and that most of the deviation from being a perfect predictor is likely coming from another source --- e.g., nonlinear terms or fluctuating initial flow.

Finally, we test Gubser's estimator, 2nd line of Eq.~(\ref{GY}). This particular quantity gives less weight to the periphery. This makes the estimator much worse for central collisions, as expected from the discussion above, but it is also worse at all other centralities.

Overall, $\varepsilon_2$ calculated with energy or entropy density weighting is a very good predictor of $v_2$, while other quantities that are significantly more sensitive to the periphery compared to the center of the system, or vice versa, are worse.
\subsection{Triangular flow}
\begin{figure}
\includegraphics[width=\linewidth]{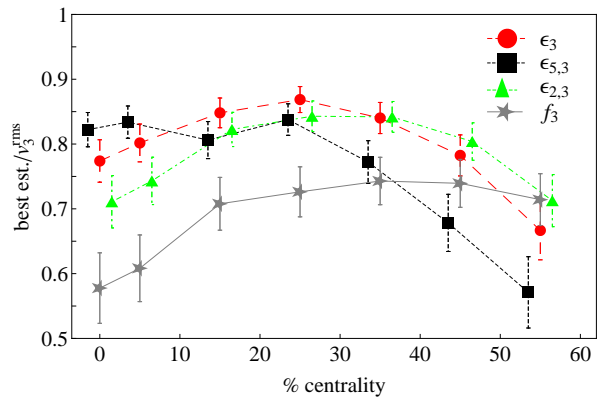}
\caption{(Color online)Best estimator for triangular flow divided by rms $v_3$ for various definitions of $\varepsilon_3$. The leftmost points correspond to 115 events with exactly zero impact parameter.  Error bars represent statistical uncertainty from the finite number of events.  Triangles (squares) have been given an x-offset of 1.5 (-1.5) for readability.}
\label{fig:v3}
\end{figure}
Similar to elliptic flow and eccentricity, triangular flow $v_3$ is thought to be a hydrodynamic response to an initial ``triangularity'' in the initial state:
to define the triangularity, Alver and Roland~\cite{Alver:2010gr} originally suggested $\varepsilon_{2,3}$
(following the notation of Eq.~(\ref{epsmn})).
However, the numerator cannot be simply expressed in terms of the moments (\ref{moments}).
Recently it has been more common to use $\varepsilon_{3,3}$~\cite{Petersen:2010cw}, which we denote by
$\varepsilon_3$.
In this case, the numerator is $W_{3,3}/W_{0,0}$, but the denominator $\{r^3\}$ is not a simple moment.
One could instead replace $\{r^3\}$ with a power of the lowest moment $\{r^2\}^{3/2}$.
This gives an almost indistinguishable result in our analysis, so we only show the curve for the ``standard'' denominator.\footnote{Values of $\varepsilon_3$ are slightly larger with $\{r^2\}^{3/2}$ than with $\{r^3\}$ denominator, and $\varepsilon_3$ is no longer bounded by 1,
but this is almost exactly compensated by the smaller value of $k$ from Eq.~(\ref{bestk}).}

Figure~\ref{fig:v3} shows the ratio of the rms value of the best estimator to the rms value of the calculated $v_3$.   The triangularity $\varepsilon_{2,3}$ is a worse predictor of $v_3$ than $\varepsilon_3$ below 30\% centrality, but slightly better above 40\%.  As with $v_2$, the second-lowest moment $\varepsilon_{5,3}$ is a slightly better predictor for central collisions, but worse for non-central collisions, while $\varepsilon_{7,3}$ (not shown) is worse everywhere.  This again signals a somewhat stronger sensitivity to the periphery of the collision region in central collisions than in peripheral collisions, though again the moment $f_3$, which has a strong sensitivity to the interior, is worse at all centralities.

As with elliptic flow, replacing an energy density weight with an entropy density weight in the calculation of $\varepsilon_3$ gives results (not shown) that are slightly better for central collisions and worse for peripheral collisions, but that are essentially equivalent.

Finally, using a sum of the lowest two moments $\varepsilon_3$ and $\varepsilon_{5,3}$ (not shown) reproduces the highest points on the figure.  I.e., it shows no improvement over the term that is individually the best predictor, except above 40\% centrality, where it follows the $\varepsilon_{2,3}$ result.

Thus, $\varepsilon_3$ is a very good predictor, with slightly too little sensitivity to the periphery in central collisions and slightly too much in peripheral collisions, but quantities with too-different $r$-dependence are everywhere worse.

\subsection{Quadrangular flow}
\begin{figure}
\includegraphics[width=\linewidth]{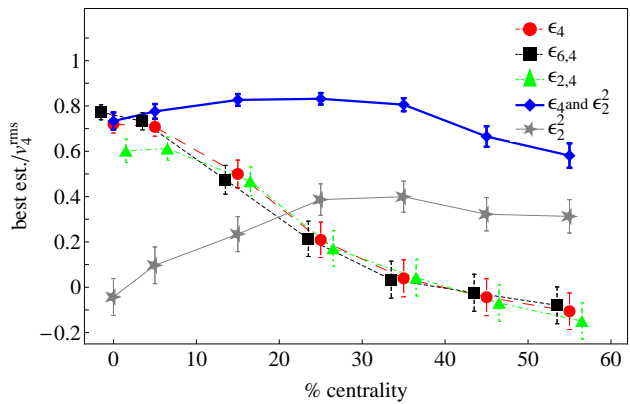}
\caption{(Color online) Best estimator for quadrangular flow divided by rms $v_4$ for various choices of the estimator.  Negative values indicate an anticorrelation with the estimator. The leftmost points correspond to 115 events with exactly zero impact parameter.  Error bars represent statistical uncertainty from the finite number of events.  Triangles (squares) have been given an x-offset of 1.5 (-1.5) for readability.}
\label{fig:v4}
\end{figure}
Like the triangularity, the quadrangularity has inconsistent definition in the literature.  Some authors use $\varepsilon_{2,4}$~\cite{Alver:2010dn},
others use $\varepsilon_{4,4}$ (which we denote by $\varepsilon_4$)~\cite{Lacey:2010hw}. Another possible choice is, in the latter case, to replace $\{r^4\}$ with $\{r^2\}^2$ in the denominator.  In this case, the estimator is equally good, and so the results are not shown with the rest of the results in Fig.~\ref{fig:v4}.

These results differ qualitatively from our results for $v_2$ and $v_3$. For $v_2$ and $v_3$, all the estimators we have tested give good results, for all centralities.
By contrast,  $\varepsilon_4$ gives reasonable predictions  only for central collisions. The agreement becomes much worse for peripheral collisions, in agreement with previous analysis~\cite{Qiu:2011iv}. In fact, $v_4$ is anticorrelated with $\varepsilon_4$ for the most peripheral points.
Despite suggested scaling similarities between the centrality dependence of $\varepsilon_4$ and $v_4$~\cite{Lacey:2010hw}, this shows that $\varepsilon_4$ cannot be used as an estimator of $v_4$ on an event-by-event basis for non-central collisions.

When using an $r^2$ weight ($\varepsilon_{2,4}$), the result is significantly worse for central collisions.  Using the next highest moment $\varepsilon_{6,4}$ is slightly better for central collisions, but all higher moments are worse.

For peripheral collisions, the asymmetry of the nuclear overlap
region, causes $\varepsilon_2$ to be significantly larger than other
moments such as $\varepsilon_4$.  This raises the possibility that
non-linear terms involving $\varepsilon_2$ may be important.  The
first such term allowed by symmetry is proportional to
$\varepsilon_2^2$ --- that is, $k(\varepsilon_2  {\rm e}^{2i\Phi_2})^2$~\cite{Luzum:2011mm}.  Fig.~\ref{fig:v4} shows that, indeed, this term alone provides a reasonable estimator for non-central collisions.  More interestingly, including both terms, i.e.,
\begin{equation}
v_4 {\rm e}^{4i\Psi_4}=k\varepsilon_4  {\rm e}^{4i\Phi_4}+k'\varepsilon_2^2  {\rm e}^{4i\Phi_2},
\end{equation}
results in an excellent predictor for all centralities.
For central collisions, $\varepsilon_4$ and $\varepsilon_2$ are both small and of the same order of magnitude, so that the first term dominates.
For all other centralities, both terms are of comparable magnitudes, and the combination gives a much better result than either individual term.

Note that the rms $v_4$ was first measured in 2011~\cite{Adare:2011tg,Aamodt:2011by,Jia:2011hf,Li:2011mp}.
In earlier analyses,  $v_4$ was determined with respect to the event plane from elliptic flow~\cite{Adams:2003zg,Adare:2010ux}.
The measured quantity is then $\langle v_4{\rm e}^{4i\Psi_4}v_2^2{\rm e}^{-4i\Psi_2}\rangle/\langle v_2^2\rangle$~\cite{Gombeaud:2009ye},
not the rms $v_4$.

\subsection{Pentagonal flow}
\begin{figure}
\includegraphics[width=\linewidth]{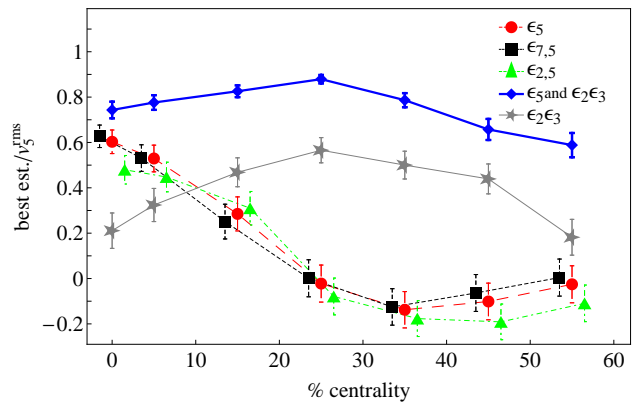}
\caption{(Color online) Best estimator for pentagonal flow divided by rms $v_5$ for various choices of the estimator.  Negative values indicate an anticorrelation with the estimator. The leftmost points correspond to 115 events with exactly zero impact parameter.  Error bars represent statistical uncertainty from the finite number of events.  Triangles (squares) have been given an x-offset of 1.5 (-1.5) for readability.}
\label{fig:v5}
\end{figure}
Figure~\ref{fig:v5} shows the results for $v_5$, which are very similar to the results for $v_4$.  Here, the non-linear term $\varepsilon_2\varepsilon_3$ becomes important even for central collisions.  The predictor
\begin{equation}
v_5 {\rm e}^{5i\Psi_5}=k\varepsilon_5  {\rm e}^{5i\Phi_5}+k'\varepsilon_2{\rm e}^{2i\Phi_2}\varepsilon_3  {\rm e}^{3i\Phi_3},
\end{equation}
does an excellent job at all centralities.
%
%
%
%
\section{Conclusions}

In this work, we have quantitatively tested to what extent anisotropic flow can be predicted from the initial density profile in event-by-event ideal hydrodynamics with realistic initial conditions.
We have shown that the participant eccentricity $\varepsilon_2$ gives a very good prediction of elliptic flow for all centralities.
We have also shown that the definition of $\varepsilon_3$ with $r^3$ weights \cite{Teaney:2010vd,Petersen:2010cw}
gives a better prediction of triangular flow than the previous definition with $r^2$ weights.
Gubser's moments~\cite{Gubser:2010ui} give worse results for both $v_2$ and $v_3$.
Higher harmonics $v_4$ and $v_5$ can be well predicted from the corresponding eccentricities $\varepsilon_4$ and $\varepsilon_5$ (again defined with $r^4$ and $r^5$ weights rather than with $r^2$ weights) only for central collisions. For noncentral collisions, a good predictor of $v_4$ must include two terms, proportional to   $\varepsilon_4$ and  $\varepsilon_2^2$. Likewise, $v_5$ has contributions proportional to $\varepsilon_5$ and  $\varepsilon_2\varepsilon_3$.  Defining the eccentricities with energy or entropy density, or using the density at a midrapidity slice or over a finite longitudinal range is largely a matter of preference, and does not make a significant difference.

These results provide an improved understanding of the hydrodynamic response to the initial state in realistic heavy-ion collisions, and provide a more direct link between experimental data and properties of the initial stage of the collision.  This will allow for the construction of more realistic models for the early-time collision dynamics, and thus a significant reduction in the systematic uncertainties of extracted bulk properties of the system.
%
%
%
%
\begin{acknowledgments}
%
This work is funded by ``Agence Nationale de la Recherche'' under grant
ANR-08-BLAN-0093-01, by Cofecub under project Uc Ph 113/08;2007.1.875.43.9, by FAPESP under projects 09/50180-0 and 09/16860-3, and by CNPq under project  301141/2010-0. ML is supported by the European Research Council under the
Advanced Investigator Grant ERC-AD-267258.
\end{acknowledgments}

\end{document}